\documentclass[journal]{IEEEtran}

\usepackage[utf8x]{inputenc} 

\usepackage{cite}

\ifCLASSINFOpdf
   \usepackage[pdftex]{graphicx}
\else
   \usepackage[dvips]{graphicx}
\fi
\usepackage[cmex10]{amsmath}
\newcommand{\disp}{\displaystyle}
\newcommand{\ba}{\begin{array}}
\newcommand{\ea}{\end{array}}
\newcommand{\btab}{\begin{table}}
\newcommand{\etab}{\end{table}}
\newcommand{\bcen}{\begin{center}}
\newcommand{\ecen}{\end{center}}
\newcommand{\btabb}{\begin{tabular}}
\newcommand{\etabb}{\end{tabular}}
\newcommand{\bea}{\begin{eqnarray}}
\newcommand{\eea}{\end{eqnarray}}
\newcommand{\beqn}{\begin{equation}}
\newcommand{\eeqn}{\end{equation}}
\newcommand{\beqnt}{\begin{equation*}}
\newcommand{\eeqnt}{\end{equation*}}
\newcommand{\bex}{\begin{example}}
\newcommand{\eex}{\end{example}}

\newcommand{\dsuml}{\disp\sum\limits}

\newcommand{\nrxi}[1]{\mathcal{N}^{#1}}

\usepackage{xfrac}

\usepackage[list=true, justification=centering, hypcap=true]{subcaption}
\usepackage{scalerel}

\IEEEoverridecommandlockouts
\usepackage{tikz}
\usepackage{textcomp}
\usepackage{lipsum}

\newcommand\copyrighttext{%
	\footnotesize \textcopyright 2016 IEEE. Personal use of this material is permitted.
	Permission from IEEE must be obtained for all other uses, in any current or future
	media, including reprinting/republishing this material for advertising or promotional
	purposes, creating new collective works, for resale or redistribution to servers or
	lists, or reuse of any copyrighted component of this work in other works. %
	}
\newcommand\copyrightnotice{%
	\begin{tikzpicture}[remember picture,overlay]
	\node[anchor=south,yshift=8pt] at (current page.south) {\fbox{\parbox{\dimexpr\textwidth-\fboxsep-\fboxrule\relax}{\copyrighttext}}};
	\end{tikzpicture}%
}

\begin{document}
%
\title{ISI-Aware Modeling and Achievable Rate \\Analysis of the Diffusion Channel}
%
%
%


\author{Gaye~Genc,
		Yunus~Emre~Kara,
        H.~Birkan~Yilmaz,~\IEEEmembership{Member,~IEEE}
        and Tuna~Tugcu,~\IEEEmembership{Member,~IEEE}
\thanks{G. Genc, Y. E. Kara, H. B. Yilmaz, and T. Tugcu are with the Bogazici University Department of Computer Engineering, 34342, Bebek, Istanbul, Turkey
 E-mails: \{gaye.genc, yunus.kara, birkan.yilmaz, tugcu\}@boun.edu.tr}
 \thanks{This project is partially supported by the State Planning Organization (DPT) of the Republic of Turkey under the project TAM with the project number 2007K120610 and Bogazici University Research Fund (BAP) under grant number 7436.}}

%
%

\markboth{IEEE COMMUNICATIONS LETTERS}%
{Author 1 \MakeLowercase{\textit{et al.}}: ISI-Aware Modeling and Achievable Rate Analysis of the Diffusion Channel}
%



\maketitle

\copyrightnotice

\begin{abstract}
Analyzing the achievable rate of molecular communication via diffusion (MCvD) inherits intricacies due to its nature: MCvD channel has memory, and the heavy tail of the signal causes inter symbol interference (ISI). Therefore, using Shannon's channel capacity formulation for memoryless channel is not appropriate for the MCvD channel. Instead, a more general achievable rate formulation and system model must be considered to make this analysis accurately. In this letter, we propose an effective ISI-aware MCvD modeling technique in 3-D medium and properly analyze the achievable rate. 
\end{abstract}


%
\IEEEpeerreviewmaketitle

\section{Introduction}
\IEEEPARstart{M}{olecular} communication is a new interdisciplinary research paradigm in the nanonetworking domain that is related to nanotechnology, biotechnology, and communication technology \cite{nakano2013molecularC}. Molecular communication via diffusion (MCvD) is an effective method since it does not require any infrastructure, and the propagation of the molecules by free diffusion is energy efficient \cite{kuran2010energy}. In MCvD, a number of micro and nano machines residing in a fluid environment communicate through molecules that are discharged into the communication medium. Following the physical characteristics of the channel, these molecules propagate through the environment via diffusion. Some of these molecules arrive at the receiver (i.e. hit the receiver) and form chemical bonds with the ligand receptors on the surface of the receiver. The properties of these received molecules (e.g., concentration, type) constitute the received signal.

The received molecular signal has a heavy tail, hence the molecules arriving in the subsequent symbol slots cause significant inter symbol interference (ISI). Therefore, MCvD channel is an ISI channel with memory, and the achievable rate of the MCvD channel must be analyzed accordingly. In the literature, channel capacity evaluations are carried out under more simple cases and assumptions.  In~\cite{kuran2010energy,liu2015channel,atakan2007anInformationTA}, the channel capacity is investigated by utilizing Shannon's formulation that is for memoryless channels, but the MCvD channel is not memoryless. In~\cite{atakan2007anInformationTA}, channel capacity in a 3-D medium is considered, however, the reception process does not consider the first passage. The reception process without first passage is easier to model but it deviates receiver dynamics from reality. For most of the cases observed in the nature, almost all of the molecules contribute to the signal only once~\cite{cuatrecasas1974membraneR}. In~\cite{pierobon2013capacityOD}, the authors derive a closed-form expression for the lower bound on channel capacity of diffusion-based molecular communication in gaseous environments. In~\cite{meng2012diffusionBB}, the diffusion-based molecular communication channel is considered with on-off keying modulation and the authors present formulations for the optimal decision threshold and the  mutual information of the diffusion channel. However, in both of the works, the derivations are based only on the diffusion process, without considering absorption/reception (i.e., the molecules can enter and exit the receiver multiple times without any resistance or change in the movement as if the receiver is transparent to the molecules' movement). In \cite{srinivas2012molecular}, information theoretical aspects of the diffusion channel is investigated in 1-dimensional (1-D) environment. In~\cite{nakano2012channel}, channel capacity formulations are derived when the messenger molecules degrade during diffusion in the medium. Although they provide a solid formulation for the channel capacity, they only consider 1-D environment and analyze a few performance metrics. 

The contributions of this paper are twofold; one is introducing a more realistic analytical model for the MCvD system surpassing the aforementioned shortcomings in the literature, by taking the memory of the channel (i.e. ISI effect) into consideration. The second is using this analytical model to introduce a stricter upper bound on the MCvD communication by employing the achievable rate calculations for finite state ISI channels. We start by defining the MCvD system and proposing our model. Then, we statistically verify the validity of the proposed model by extensive simulations with goodness of fit analysis. We conclude by presenting the realistic upper bounds on the MCvD achievable rate using our verified model.

\section{Modeling the Molecular Communications Channel Considering the ISI}
\label{sec:model}
We model a communication system composed of a fluid environment and
a pair of spherical devices, each called a Nanonetworking-enabled Node (NeN);
one as the transmitter and the other as the receiver. In MCvD, the
information is transmitted between the transmitter and the receiver
through the propagation of certain molecules via diffusion
\cite{nakano2013molecularC}. These molecules are called the Messenger
Molecules (MMs). The MMs diffuse throughout a drift-free environment obeying the laws of Brownian motion. We do not consider the
collisions between MMs for the sake of simplicity. In this paper we focus on Binary Concentration Shift Keying
(BCSK) in which 1-bit information is sent in each symbol duration \cite{kuran2010energy}.

The messenger molecules, the transmitter, and the receiver are assumed to have
spherical bodies. Whenever an MM's body coincides with the body of the
receiver, the MM is assumed to be received and removed from the environment.
A single MM reaches the receiver before a given
deadline with a certain probability. This probability, known as the \emph{first passage probability}, is affected mainly by the diffusion coefficient, transmitter-receiver properties, and the distance between the transmitter and the receiver,  which is denoted by $d$. In the literature, reception process considering first passage probability is mostly examined in a 1-D environment~\cite{nakano2012channel, srinivas2012molecular}. First passage probability in a 3-D environment with a point source is formulated in~\cite{yilmaz2014_3dChannelCF}. In our setup the transmitter node is a spherical body, which is more realistic and complex.

We assume that time is divided into equal sized time slots of length $t_s$, called the symbol duration. At the start of a symbol slot, $s$, the transmitter NeN releases a predefined number of molecules, $\mathcal{N}_s$, to the communication environment, where $\mathcal{N}_s$ depends only on the transmitted bit $x_s$. In our channel model, the transmitted bits are independent and identically distributed. 

Arising from the probabilistic dynamics of Brownian motion, the MMs move randomly and reach the receiver at different symbol slots. Note that it is possible for an MM to miss the receiver since the first passage process is not recurrent in 3-D environment~\cite{yilmaz2014_3dChannelCF}, i.e., molecules have positive survival probability.

Let us denote the probability of an MM being received with a delay of $i$ symbols as $p_i$. The reception event of an MM can be modeled as a Bernoulli trial with success probability $p_i$, and therefore the reception event of $\mathcal{N}_s$ MMs can be modeled with the Binomial distribution. We denote the number of molecules that are emitted at the beginning of the $s^{th}$ symbol slot and received during the $r^{th}$ symbol slot as the random variable $N_{s}^{r}$. We start by defining $N_{s}^{s}$, the number of MMs that experience $i\!=\!0$ symbols of delay, as the binomial random variable
\begin{equation}
	N_{s}^{s}\sim \mathcal{B}(\mathcal{N}_s,p_{0}).
\end{equation}

The case is not simple for the MMs that are not released and received in the same symbol slot. These MMs cause ISI in the channel and modeling the number of such MMs is the main contributing factor in our channel model. The number of MMs that experience a delay of $i\!>\!0$ symbols depends on the number of MMs that had been released in the same symbol slot and experienced a delay less than $i$ symbols (i.e. $N_{s}^{s+i}$ depends on $N_s^{s+j}$, $\forall j \in \{0,\dots,i-1\}$). We define this conditional distribution as
\begin{equation}
N_{s}^{s+i}\stretchrel{\mid}{\left(N_{s}^{s},\dots,N_{s}^{s+i-1}\right)} \sim \mathcal{B}\left(\mathcal{N}_s-\sum_{j=0}^{i-1}N_{s}^{s+j},p_{i}^{*}\right)
\end{equation}
where $p_{i}^{*}$ is the success probability of a remaining MM being received $i$ symbols later, given that it was not received before. $p_{i}^{*}$ is calculated using the previously defined $p_{i}$ values as
\begin{equation}
	p_{i}^{*} = \dfrac{p_i}{1-\sum_{j=0}^{i-1}p_j}.
\end{equation}

At the receiver side, the total number molecules received in the $r^{th}$ symbol slot ($\nrxi{r}$) not only depends on the number of MMs released in the $r^{th}$ symbol slot, but also on the previous transmissions. 
Since $p_i$ values are exponentially decaying \cite{kuran2010energy}, expected $N_{s}^{s+i}$ also decreases rapidly as $i$ increases. Therefore, there exists an integer $\eta\!\geq\!0$ for which $N_{s}^{s+\eta+i}$ values are negligible. Taking this fact into account, a transmission is only significantly affected by the $\eta$ previous symbols before itself. We call this $\eta$ value as the \textit{ISI-awareness window} of the proposed model, which is the minimum number of channel coefficients required for realistic modeling. Then, $\nrxi{r}$ can be represented as the sum of $\eta+1$ independent binomially distributed random variables
\begin{equation}
\nrxi{r} = \dsuml_{s=r-\eta}^{r} N_{s}^{r} = \dsuml_{k=0}^{\eta} N_{r-k}^{r}.
\end{equation}
Thus, the demodulated bit $y_r$ by the receiver NeN at the $r^{th}$ symbol can be computed as $y_r = 0$ if $\nrxi{r} \leq \tau$, $y_r = 1$ if $\nrxi{r} > \tau$, where $\tau$ is the symbol demodulation threshold. Note that, $y_r$ depends on the previous emissions (previous $x_s$ values), and we take $\eta$ previous symbols into account for modeling the impact of ISI. Therefore, we have
\begin{align}
p(y_r\!=\!0|x_{1},\dots,x_r) &\!=\! p(\nrxi{r} \leq \tau|x_{r-\eta+1},\dots,x_r)\nonumber\\
&\!=\! \sum\limits_{i=0}^\tau p(\nrxi{r}=i|x_{r-\eta+1},\dots,x_r)\label{eq:prob_sum}\\
&\!=\! \sum\limits_{i=0}^\tau p\!\left(\dsuml_{k=0}^{\eta}N_{r-k}^{r}\!=\!i \middle\vert x_{r-\eta+1},\dots,x_r\right).\nonumber
\end{align}
Calculating the probability in \eqref{eq:prob_sum} is done by $\eta$ convolutions since $\nrxi{r}$ is the sum of $\eta+1$ random variables.

In the literature, it is considered that the most significant effect is due to 1-previous symbol (\!\cite{kuran2010energy, arjmandi2013diffusionBN}), which corresponds to having $\eta\!=\!1$. In Section~\ref{sec:model_verification}, we employ an analysis focusing on the deviation from the simulation results for validating our model and determine the required value for $\eta$.

\section{Verification of the ISI-aware MCvD model}
\label{sec:model_verification}

In order to ensure the validity of our proposed analytical model, we implement a particle tracking based MCvD simulator where the movement of each MM is monitored. We set up a simulation environment mimicking the pancreatic islets where the transmitter and receiver NeNs are similar in size to the pancreatic beta cells ($10 \,\mu m$ radius), MMs are similar to the insulin hormone ($2.5\, nm$ radius), and the typical distance between the communicating pair being one cell at most ($4\!\sim\!24 \,\mu m$) \cite{alberts2010molecular}. The diffusion coefficient is $79.4 \,\mu m^2/sec$, which is in accordance with the insulin hormone. The symbol duration $t_s$ is selected as $0.1\left(\sfrac{d}{2}\right)^2 sec$ as per the setting in \cite{kuran2010energy}. This choice is also supported by the fact that the propagation delay of MCvD is shown to increase by a factor of $d^2$ in \cite{yilmaz2014_3dChannelCF}, which is drastically different than the EM spectrum communication. The number of molecules released each symbol, $\mathcal{N}_s$, is $0$ for the bit $0$ and $50$ for the bit $1$. Therefore, the demodulation threshold $\tau$ is in the range $1 \!\sim\! 50$. We investigate the ISI-awareness window $\eta$ for the range $0\!\sim\!14$. Each simulation consists of a random $10\,000$-bits long transmission, repeated $30$ times. To investigate whether our proposed model fits the simulation results, we conduct a goodness of fit analysis using Pearson's $\chi^2$-test. The test compares the probabilities $p(y_r|x_{r-\eta+1},\dots,x_r)$ obtained from the simulations with those derived from our proposed model.

\begin{figure}[htb]
	\centering
	 \begin{subfigure}[t]{.39\columnwidth}
		    	\centering
		    	\includegraphics[width=\columnwidth]{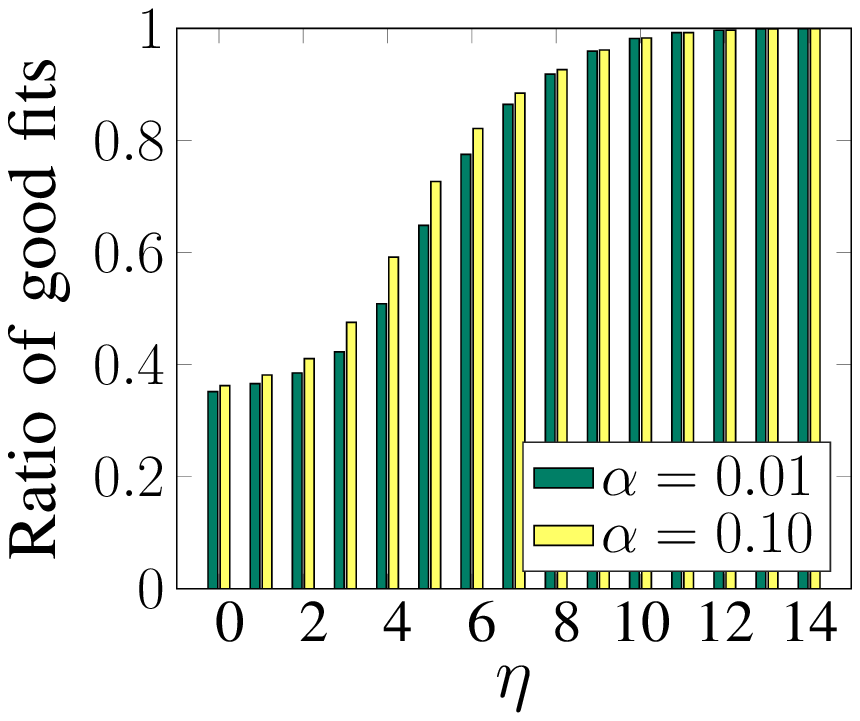}
		    	\captionsetup{skip=-2pt}
		    	\caption{ }\label{fig:goodness_vs_H}
	 \end{subfigure}
	 \begin{subfigure}[t]{.49\columnwidth}
 		    	\centering
 		    	\includegraphics[width=\columnwidth]{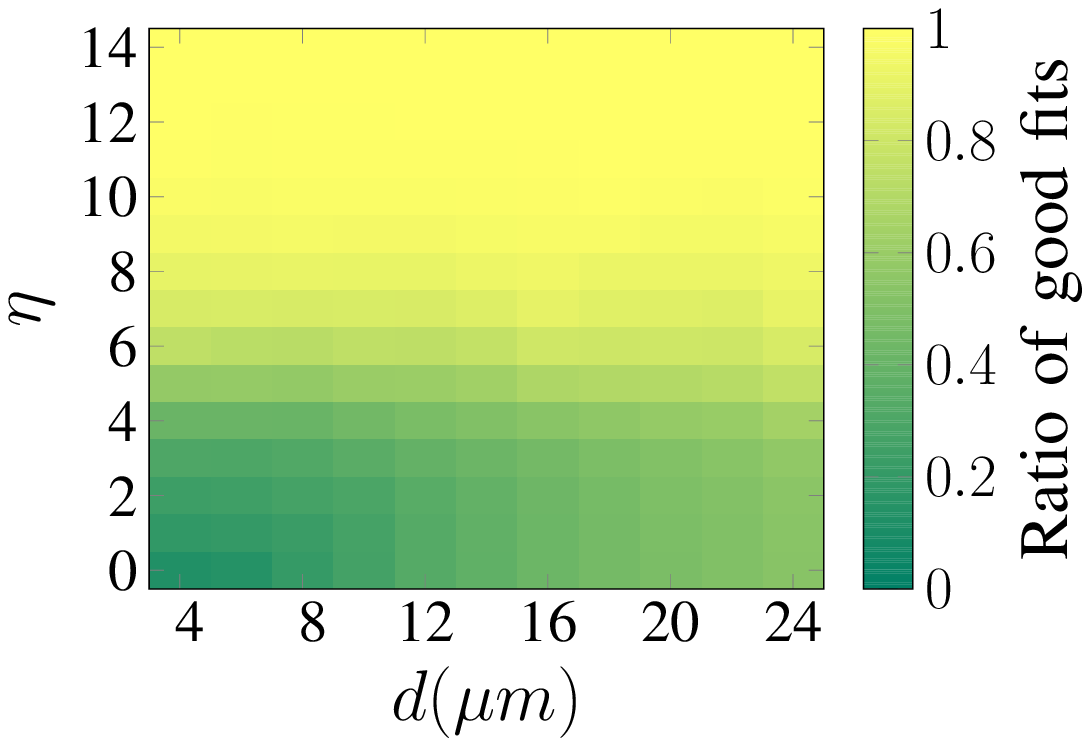}
 		    	\captionsetup{skip=-2pt}
 		    	\caption{ }\label{fig:goodness_d_vs_H}
	 \end{subfigure}	
	\caption{Model verification using Pearson's $\chi^2$ test. \subref{fig:goodness_vs_H} The change in the ratio of good fits with respect to $\eta$. $\alpha$ denotes the significance level. \subref{fig:goodness_d_vs_H} The change in the ratio of good fits with respect to $\eta$ and distance where $\alpha\!=\! 0.01$.}
\end{figure}

Figure~\ref{fig:goodness_vs_H} shows the change in the ratio of good fits with respect to $\eta$. The $\alpha$ values indicate the significance level of the Pearson's $\chi^2$ goodness test where a smaller value stands for a stricter test. The model's chance of obtaining a successful fit over all parameter combinations increases as $\eta$ is increased. We observe that the model achieves more than $95\%$ good fits when $\eta\!>\!9$ and exceeds $99\%$ when $\eta\!>\!11$.

Figure~\ref{fig:goodness_d_vs_H} shows the ratio of good fits with respect to $\eta$ and source-receiver NeN separation. A curious trend is observable for shorter distances, where decreasing $\eta$ fails in terms of model fit much rapidly as opposed to farther distances. This is due to the fact that although shorter distances have a higher probability of MM reception, the ISI is also stronger since many lagging molecules are absorbed before they have a chance to scatter away.

\begin{figure}[htb]
	\centering
	\includegraphics[width=0.8\columnwidth]{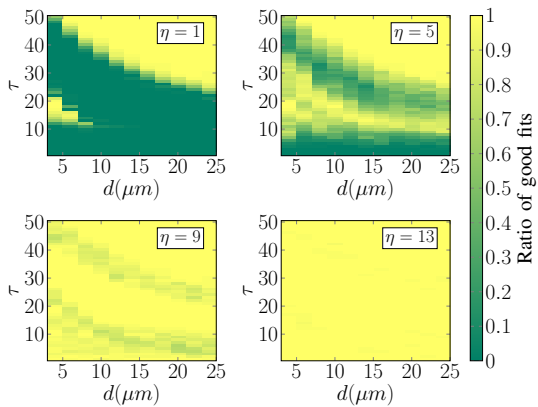}
	\caption{The change in the ratio of good fits with respect to the demodulation threshold $\tau$ and distance $d$. Good fits are calculated using the $\chi^2$-test with a significance level of $0.01$. The brighter intensity indicates the ratio of the scenarios where the model successfully fits the simulation results. When ${\eta\!=\!13}$, the model is able to fit the simulation results for almost all deployment scenarios over all possible $\tau$ values.}
	\label{fig:goodness_tau_vs_d}
\end{figure}

In Figure~\ref{fig:goodness_tau_vs_d}, heat maps of ratio of good fits for four different $\eta$ values are presented with details in terms of distance and threshold values. The brighter areas show the parameters for better agreement with the model. There are four main areas of interest, which are clearly observable in the $\eta\!=\!5$ case. Upper right portions of the graphs indicate that simulation and the model results overlap since larger $\tau$ values are biased towards demodulating bits as 0 in both the model and the simulations. The bright intensity patch towards the middle corresponds to the cases where the narrow band of  appropriate demodulation thresholds result in overlapping results for the simulation and model. The two lower intensity bands, which fade away with increasing $\eta$, correspond to the disagreement scenarios. The lower band represents the cases where the ISI easily causes incorrect demodulations due to low $\tau$ selection. The upper low intensity band lying between the high-intensity zones depicts the cases where the model expects incorrect demodulation of sent bits, mostly for 1, due to higher $\tau$ selection. Non-accurate modeling occurs due to disregarding or imprecise handling the ISI. Our aim is to achieve a robust model where the simulations and model results fit regardless of the selection of $\tau$. Increasing $\eta$ incorporates the ISI effect precisely into the model and the model fits the simulations regardless of the demodulation threshold, which is the case for $\eta\!=\!13$. These suitable $\eta$ values might change for MCvD setups other than those presented here; however, we see that the literature standard of $\eta=1$ is overly optimistic. Note that proposing a globally suitable $\eta$ is beyond the scope of this work.

\section{Achievable Rate Analysis of the Channel}
\label{sec:achievable_rate}
Current works in the literature use Shannon's classical channel capacity formula $C\! =\! \sup\limits_{p(x)} I(X;Y)$ for analyzing the achievable rate of the MCvD channel, since it provides a convenient closed-form solution. However, this formula is only valid for memoryless channels, and by having ISI, the MCvD channel is not a memoryless one. Thus, we need to employ the general formulation for the mutual information rate
\begin{equation}
I(\mathcal{X};\mathcal{Y}) = \lim\limits_{n \to \infty}\dfrac{1}{n} I(X_1,\dots,X_n;Y_1,\dots,Y_n).
\label{eq:mutinf}
\end{equation}

However, there is no closed-form solution for \eqref{eq:mutinf} in MCvD. Fortunately, the quantity can be computed numerically. We know that $I(\mathcal{X};\mathcal{Y}) = H(\mathcal{Y})-H(\mathcal{Y}|\mathcal{X})$. The quantity $H(\mathcal{Y}|\mathcal{X})$ can be easily calculated, whereas $H(\mathcal{Y})$ cannot. The Shannon-McMillan-Breiman theorem states that the sample entropy rate $\hat{H}_n(\mathcal{Y})$ converges to the true entropy rate $H(\mathcal{Y})$ with probability 1 for stationary ergodic random processes \cite[p. 644]{cover2006elements}. That is, 
\begin{equation}
\lim\limits_{n \to \infty}\hat{H}_n(\mathcal{Y})=\lim\limits_{n \to \infty}-\dfrac{1}{n} log\, p(Y_1,\dots,Y_n)=H(\mathcal{Y}).
\end{equation}
 
Since the MCvD channel is ergodic \cite{hsieh2013mathematical}, we can calculate $\hat{H}_n(\mathcal{Y})$ and estimate $H(\mathcal{Y})$ numerically by generating a long sequence of the demodulated bits $y_1,\dots,y_n$. The achievable information rate of the ergodic finite state ISI channels is a previously studied problem in the literature \cite{pfister2001achievable}. We have shown in Section~\ref{sec:model_verification} that the MCvD channel can be modeled correctly with a finite memory. Thus, we may represent it as a finite state machine by defining the states as the sequence of the last transmitted $\eta+1$ bits, resulting in $2^{\eta +1}$ states. We use the recursive calculations shown in \cite{pfister2001achievable}, by incorporating the outputs of our analytical model, namely, the demodulation probabilities presented in \eqref{eq:prob_sum}. We omit the tedious algebra required for the calculations due to page limit.

\begin{figure*}[htb]
	\centering
	\includegraphics[width=0.7\textwidth]{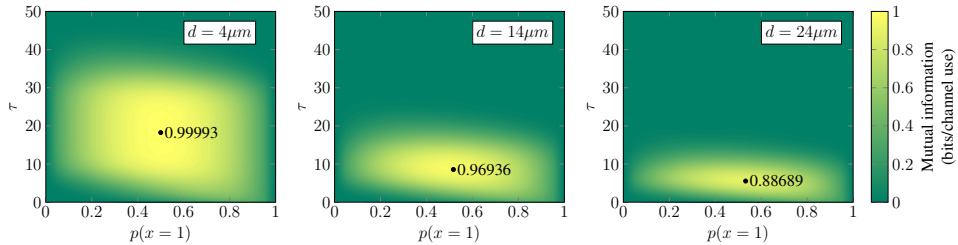}
	\caption{The change of mutual information with respect to the demodulation threshold $\tau$ and input probability distribution for three different NeN deployment scenarios. The mutual information is represented in bits/channel use, where the annotated points denote the achievable rate for the particular scenario.}
	\label{fig:mutinf_heatmap}
\end{figure*}
We use our verified model for achievable rate calculations of the MCvD system. Thankfully, we do not need to run highly time-consuming simulations, where the execution time depends drastically on MCvD system parameters. In Figure~\ref{fig:mutinf_heatmap}, the change in mutual information depending on the input distribution and demodulation threshold $\tau$ is given for different source-receiver separation values.  We observe that the achievable rate is obtained around the equiprobable input distribution for all distances, but the demodulation threshold varies significantly. For closer distances, the system is more robust in terms of $\tau$ selection. A greater percentage of the MMs are received at shorter distances, thus letting the receiver NeN to easily make the distinction between the bits 0 and 1. At farther distances, the achievable rate starts to decrease, and the effective operation range of $\tau$ narrows significantly. 

Notice the slight obliqueness in Figure~\ref{fig:mutinf_heatmap} for $d\!=\!4\mu m$. When a lower $\tau$ is selected, the system is inclined to demodulate more bits as $1$. In the case where this fact is supplemented with an input distribution biased towards producing more $1$s than $0$s, the probability of successful demodulation, and thus the mutual information, increases. The converse also holds for the combination of high $\tau$ and $0$-biased input distribution, which explains the obliqueness.

The observations on the achievable rate are extended in Table~\ref{tbl:achievable_rate}. When the literature standard of $\eta\!=\!1$ is selected, the achievable rate turns out to be overly optimistic, especially for increasing $d$. The realistic achievable rate of the molecular communication channel decreases slightly with distance in terms of bits per channel use. However, in order to keep the achievable rate (bits/channel use) high, the symbol duration has to increase for increasing distances. This results in a rapidly decreasing achievable rate in terms of bits per second and presents an open issue on symbol duration optimization.

\begin{table}[htb]
	\centering
	\caption{Achievable Rate of the MCvD channel using optimistic ($\eta\!=\!1$) vs. realistic ($\eta\!=\!14$) modeling}.
	\label{tbl:achievable_rate}
	\begin{tabular}{c||c||c|c|c}	
		$d$ & $t_s$& Achievable Rate ($\eta\!=\!1$)                 & \multicolumn{2}{|c}{Achievable Rate $(\eta\!=\!14$)} \\ \cline{3-5}
		$(\mu m)$ & (sec) & bits/ch. use & bits/ch. use & bits/sec        \\ \hline
		4     &  0.4 &   1.0000    &   0.9999    & 2.4998 \\
		8     &  1.6 &   0.9995    &   0.9965    & 0.6228 \\
		12    &  3.6 &   0.9953    &   0.9831    & 0.2731 \\
		16    &  6.4 &   0.9830   &   0.9540    & 0.1491 \\
		20    & 10.0 &   0.9664    &   0.9218    & 0.0922 \\
		24    & 14.4 &   0.9434    &   0.8869    & 0.0616
	\end{tabular}
\end{table}

\section{Conclusion}
In this work, we introduce a realistic analytical ISI-aware channel model for MCvD with a spherical transmitter-receiver pair in 3-D environment. We show that the MCvD channel suffers from high ISI and propose a finite-state channel model to represent the ISI effect realistically. We validate the proposed model by rigorous testing based on statistical methods and show that the assumption of a 1-symbol ISI awareness window in the literature is overly optimistic and falls short of modeling the channel correctly. Furthermore, we present a general view of the achievable rate of the channel. The achievable rate analysis shows that incorrect modeling of the channel leads to an erroneously high achievable rate. Moreover we observe that the system suffers from transmitter-receiver separation significantly under farther distances and further investigation is required for the optimal symbol duration. 

\ifCLASSOPTIONcaptionsoff
  \newpage
\fi



%

\bibliographystyle{IEEEtran}
\bibliography{references}

%

%
%
%




\end{document}